  \documentstyle[12pt]{article}

\catcode`@=11
\def\dif{{\rm d}}
\def\deriv{\@ifnextchar[{\@deriv}{\@deriv[]}}
\def\@deriv[#1]#2#3{\mathchoice%
{{\dif^{#1}#2\over\dif{#3}^{#1}}}{{\dif^{#1}#2/\dif{#3}^{#1}}}%
{{\dif^{#1}#2\over\dif{#3}^{#1}}}{{\dif^{#1}#2/\dif{#3}^{#1}}}}

\def\presup#1{{}^{#1}\kern-.15em\relax}      
\def\presub#1{{}_{#1}\kern-.12em\relax}      

\def\secteqno{\@addtoreset{equation}{section}%
\def\theequation{\thesection.\arabic{equation}}}
\def\endsecteqno{\def\theequation{\@ifundefined{chapter}%
{\arabic{equation}}{\thechapter.\arabic{equation}}}}
\newcounter{subequation}
\def\thesubequation{\alph{subequation}}
\def\sneqnarray{\stepcounter{equation}\let\@currentlabel=\theequation
\setcounter{subequation}{1}
\def\@eqnnum{{\rm (\theequation\thesubequation)}}
\global\@eqcnt\z@\tabskip\@centering\let\\=\@eqncr\let\@@eqncr=\@@sneqncr
$$\halign to \displaywidth\bgroup\@eqnsel\hskip\@centering
 $\displaystyle\tabskip\z@{##}$&\global\@eqcnt\@ne
 \hskip 2\arraycolsep \hfil${##}$\hfil
 &\global\@eqcnt\tw@ \hskip 2\arraycolsep $\displaystyle\tabskip\z@{##}$\hfil
  \tabskip\@centering&\llap{##}\tabskip\z@\cr}
\def\endsneqnarray{\@@sneqncr\egroup $$\global\@ignoretrue}
\def\@@sneqncr{\let\@tempa\relax
   \ifcase\@eqcnt \def\@tempa{& & &}\or \def\@tempa{& &}
   \else \def\@tempa{&}\fi
     \@tempa \if@eqnsw\@eqnnum\stepcounter{subequation}\fi
     \global\@eqnswtrue\global\@eqcnt\z@\cr}
\def\nobiblabels{\def\@lbibitem[##1]##2{\@bibitem{##2}}}
\catcode`@=12
\secteqno
\newcommand{\be}{\begin{equation}}
\newcommand{\ee}{\end{equation}}
\newcommand{\bea}{\begin{eqnarray}}
\newcommand{\eea}{\end{eqnarray}}
\newcommand{\bref}[1]{(\ref{#1})}
\newcommand{\ep}{\epsilon}

\newcommand{\T}{\theta}
\newcommand{\vp}{\varphi}
\newcommand{\D}{\delta}
\newcommand{\A}{\alpha} \newcommand{\B}{\beta}
\newcommand{\G}{\gamma} 
\def\pa{\partial}

\newcommand{\C}[1]{{\cal #1}}

         \newcommand{\lam}{\lambda}
            
\newcommand{\r}{\rho}           \newcommand{\s}{\sigma}

\newcommand{\Th}{\Theta}

\newcommand{\nn}{\nonumber}

\def\tA{{\tilde A}}\def\tB{{\tilde B}}
\def\tD{{\tilde D}}\def\tE{{\tilde E}}
\def\tb{{\tilde b}}

\def\NAT{non-anomalous transformation }
\def\ATs{anomalous transformations }
\def\NATs{non-anomalous transformations }
\def\Diff{diffeomorphism }
\def\CG{{\cal G}}

\def\NG{{NG }}
\font\fiverm=cmr10

\title{{\bf Nambu-Goldstone Fields, Anomalies and WZ Terms
}}
\author{{\sc J. Gomis},
        {\sc K.Kamimura$^\dagger$  and}
        {\sc R.Kuriki}$^{\dagger\dagger}$\\
\\
      \small{\fiverm{Theory Group, Department of Physics, 
                The Univ.\ of Texas at Austin}}\\
        \small{\fiverm{RLM\,5208, Austin, TEXAS and}}\\
        \small{\fiverm{Departament d'Estructura i Constituents
               de la Mat\`eria}}\\
        \small{\fiverm{Universitat de Barcelona and 
Institut de F\'{\i}sica d'Altes Energies}}\\
        \small{\fiverm{Diagonal, 647, }}
        \small{\fiverm{E-08028 BARCELONA}}\\
        \llap{$^\dagger$}%
        \small{\fiverm{Department of Physics, Toho University}}\\
        \small{\fiverm{Funabashi, }}
        \small{\fiverm{274 JAPAN}}\\
        \llap{$^{\dagger\dagger}$}%
        \small{\fiverm{Department of Physics, Tokyo Institute of Technology}}\\
        \small{\fiverm{Oh-Okayama, }}
        \small{\fiverm{Tokyo}}
        \small{\fiverm{152 JAPAN}}\\
 \small{gomis@ecm.ub.es,
kamimura@ph.sci.toho-u.ac.jp,
r-kuriki@th.phys.titech.ac.jp}\\
}
\date{}

\begin{document}

\maketitle

\thispagestyle{empty}
\begin{abstract}
We construct the Wess-Zumino terms from anomalies
in case of quasigroups for the following situations.  
One is effective gauge field theories of Nambu-Goldstone fields 
associated with spontaneously broken global symmetries
and the other is anomalous gauge theories. 
The formalism that we will develop
can be seen as a generalization of the non-linear realization 
method of Lie groups. As an example  we consider 
2d gravity with a Weyl invariant regularization 
\end{abstract}

\vskip 5mm
{\centerline {PACS 11.15.-q, 04.60.+n}}
\vskip 5mm
{\centerline {{\bf keywords} 
{non-linear realization, anomaly, Wess-Zumino term, antifield formalism}}}

\vfill
\vbox{
\hfill November 1996 \null\par
\hfill UB-ECM-PF 96/19\null\par
\hfill UTTG-18-96\null\par
\hfill TOHO-FP-9654\null\par
\hfill TIT/HEP-347
}\null
\clearpage


\section{Introduction}
\indent

Nambu-Goldstone bosons \cite{g}\cite{n}\cite{gsw} appear in the 
low energy description 
of an underlying theory with a broken global symmetry $G$. 
The Goldstone bosons, $\theta^a$, are elements of the coset $\frac GH$, 
on which the symmetry group $G$ acts in a non-linear way. 
The dynamics of the Nambu-Goldstone bosons 
can be determined from the non-linear realization 
method\cite{w}\cite{cwz}\cite{v}. 
The Lagrangian  contains  invariant terms  and terms that 
are invariant up to total derivatives \cite{wz}\cite{wi}. 
The number of quasi invariant terms is given by the
cohomology group $H^{d+1} (\frac GH, R)$\cite{hw}\cite{h}

If one introduces fictitious gauge fields\footnote{
 Here we are following the nice discussion of Weinberg in ref \cite{bookw}} 
into the underlying theory the 
global symmetry
$G$ becomes local at the classical level but is 
broken quantum mechanically 
by anomalies ${\cal A}_a(\phi_g)$, where $\phi_g$ are the fictitious 
gauge fields. 
The gauge transformations associated with $\frac GH$ becomes 
anomalous while those
associated with $H$ 
are non-anomalous.
Since the underlying theory is  anomaly free initially, 
these anomalies should be canceled
by introducing
spectator fields which are  weakly coupled  to the fictitious gauge 
fields so
 that 
the physical content of the
theory 
is unchanged. We can have in  mind, for example,  a  
$SU(3) \times SU(3)$ broken theory with massless  
trapped fermions, in this case  the spectator fields will be 
fermions that  cancel the chiral
gauge anomaly.

In the low energy description in order to have a 
vanishing anomaly \cite{th}, the gauged 
effective theory of 
Goldstone
bosons must have an anomaly associated to  the fictitious local 
symmetries. It is
equal to the one produced in the underlying theory \cite{wz}
 i.e.,
\begin{equation}
\delta_a {\cal M}_1(\phi,\theta)={\cal A}_a(\phi). 
\label{wz0}
\end{equation}
${\cal M}_1(\phi,\theta)$ gives the interaction term between the
Nambu-Goldstone fields and the fictitious gauge fields and is
called the WZ term.
It can also be used to study the interaction
of  Goldstone bosons   
coupled weakly to real gauge fields, for 
example the  Goldstone bosons
of  the standard model interacting with the weak-electromagnetic 
gauge fields.

\medskip

Another physical situation where \bref{wz0} appears is in the 
case of anomalous gauge
theories \cite{p}\cite{f}\cite{jr}. 
In these theories there are gauge degrees of freedom, $\theta$, 
that become propagating at quantum level.
They are introduced in order to cure 
the initial anomalous gauge theory. 
The dynamics of these  extra variables, 
for which there is no dynamics at classical level,
is governed by the WZ term ${\cal M}_1(\phi,\theta)$.
It differs from the case of
 broken global theories where the introduction of these fields
causes no physical change of the initial theory.

In this paper we will consider a  framework that could be 
applied
to 
study both the WZ terms for the gauged effective theories of 
Goldstone bosons
and 
anomalous gauge theories.\footnote{ From here on we call the 
ordinary Nambu-Goldstone fields 
and the extra 
variables introduced to quantize anomalous gauge theories
as Nambu-Goldstone (NG) fields.}
The construction of a dynamical action for \NG fields from the 
anomalies, 
\bref{wz0}, has been 
investigated in the literature for Yang-Mills type 
gauge theories
\cite{wz}\cite{wi}\cite{wu}\cite{msz}\cite{chz}.
For more general gauge theories with soft and/or reducible 
open gauge algebras \cite{dh}\cite{batalin}\cite{bv}\cite{review} 
we have constructed the WZ terms  using an extended field-antifield 
formalism for the case in which the anomalous transformations 
form a 
closed algebra \cite{gp1}\cite{gp2}\cite{gpz}\cite{gkpz}\cite{gkk}. 
Here we will analyze  a more general situation in which each   
anomalous and non-anomalous transformations do not form closed algebras. 
In this case we need, 
in general, to introduce \NG fields for all the 
anomalous and 
non-anomalous transformations\footnote{For YM type theories 
this possibility 
has been studied for example in \cite{wu}\cite{fss}.}. 
The presence of the \NG fields associated to the non-anomalous 
transformations will generate  gauge symmetries 
(hidden symmetries) of the WZ terms. 
Our analysis  can be seen as a generalization  of the non-linear 
realization method of
Lie groups.

As an example we consider the case of a true anomalous gauge 
theory, the
two dimensional gravity \cite{p}\cite{p1} with a Weyl 
invariant regularization \cite{kmk}\cite{j}. 
In this case the Weyl and 
the area-preserving \Diff transformations (SDiff), are 
non-anomalous. 
The anomalous transformations are other independent combination 
of the 
two dimensional diffeomorphism (Diff$_2$). 
The local splitting into anomalous 
and non-anomalous transformations is not possible  in 
this case. 
The WZ term constructed in terms of the original \NG fields 
 associated 
to Diff$_2$ is  non-local. However, in a natural way we can
find a non-local changes of coordinates to a new set of
\NG fields such that the WZ term has a local form and 
coincides with the one introduced recently in the 
literature \cite{kmk}\cite{j}.
The non-local change of coordinates will give us a \NG field  
parametrizing the coset, $\frac{Diff_2}{SDiff}$. 
\medskip

This paper is organized as follows.
In section 2, we derive a general expression of the WZ terms and 
discuss
the relation with the non-linear realization method. 
In section 3 the WZ action is constructed for the 2D gravity
in a Weyl and SDiff invariant form.
In section 4 we give a short summary and conclusions. 
There are three appendices devoted to more technical details.
\medskip

\section{   WZ terms }
\indent

In this section we will construct the antifield independent 
part of the WZ 
terms for general gauge theories with 
irreducible and closed algebras.\footnote{ 
The generalizations to reducible and on-shell theories may be
examined by consulting \cite{gkpz}.}
The gauged effective field theories of \NG bosons and 
anomalous gauge theories
have  anomalous transformations that in general do not form  
a sub-group.
For quasigroups the \NATs are not closed also in general.
In case of Lie groups the  \NG fields $\theta^a$ are elements of the coset 
$G/H$ \cite{w}\cite{cwz}\cite{v} ,
 with  $H$ being the non-anomalous sub-group, and 
 the WZ terms are constructed in 
ref.\cite{wz}\cite{wi}\cite{wu}\cite{msz}\cite{hw}\cite{h}\cite{chz}.
In more general cases we should introduce \NG fields 
associated with the 
 {\it non-anomalous} gauge transformations 
in addition to ones associated to the anomalous transformations,   
the presence of these \NG fields
 will make the WZ terms invariant under  new gauge 
transformations of \NG fields.

\subsection{Anomalous and Non-anomalous Gauge Transformations} 
\indent

Let us consider an action   
$S_0(\phi)$ with a irreducible gauge structure  ${\cal G}$,
that we  assume to be closed off shell for simplicity. 
In the case of global 
spontaneously broken theories $S_0(\phi)$ 
describes
the underlying gauge theory  of
the gauged effective field theory of \NG bosons and in the case of
 anomalous gauge theories $S_0(\phi)$ is the classical 
action.
In both cases
the effective actions $\Gamma$ are not BRST invariant 
under all classical symmetries but we have anomalies $ {\C{A}}$;
\be
\delta\Gamma= i\hbar{\C{A}}.
\label{varwz}
\ee
At one loop it has a form\footnote{
For anomalies appearing in perturbative calculations, in case of 
closed theories, we can study the antifield independent part 
separately from the antifield dependent part, see ref \cite{gp2}}
\be
 {\C{A}}~=~{\C{A}}_{\mu}(\phi) c^{\mu},
\label{anomaly}
\ee
where ${\C{A}}_{\mu}$'s are local functions of the 
classical fields $\phi^j$
and $c^\mu$'s are ghosts associated with the gauge 
transformations 
of ${\cal G}$.  
Some of them may be identically zero and some are 
linearly dependent. It 
may be expressed in 
terms of independent components $\tilde{{\C{A}}}_a$ as
\be
{\C{A}}_{\mu}(\phi) c^{\mu}= \tilde{{\C{A}}}_a(\phi)
\tilde{Z^a_{\mu}}(\phi) c^{\mu}.
\label{non1}
\ee
We can define therefore a 
subset of non-anomalous transformations 
as the ones whose infinitesimal parameters (ghosts) 
obey the system 
of local (differential) equations
\be
\tilde{Z^a_{\mu}}(\phi) \ep^{\mu}=0.
\label{system1}
\ee
The solutions are expressed as
\be
\ep^\mu~=~Z^\mu_A~\tilde\ep^A,~~~~~~{\rm with}~~~~~~
\tilde{Z^a_{\mu}}~Z^\mu_A~=~0,
\label{NAtrans}
\ee
where $\tilde\ep^A$'s are the parameters of infinitesimal \NATs. 
Here and hereafter we use the indices $a,b,...$ as the 
anomalous components,
$A,B,...$ as the 
non-anomalous ones and $\mu,\nu,...$ for all of them \footnote{ The
De Witt 
condensed sum conventions for repeated indices are also understood.}.

Since $\tilde{Z^a_{\mu}}$'s are local operators $Z^\mu_A$'s 
are chosen to be local. 
The infinitesimal anomalous transformations are ones 
that do not verify \bref{system1}. 
In general the infinitesimal 
gauge transformations could be parametrized as
\be
\ep^\mu~=~Z^\mu_a~\tilde\ep^a+Z^\mu_A~\tilde\ep^A,
\label{parameter}
\ee
here $\tilde\ep^a$'s 
are the anomalous and $\tilde\ep^A$'s are the 
non-anomalous parameters and
$(Z^\mu_a,Z^\mu_A)$ and $(\tilde Z^\mu_a,\tilde Z^\mu_A)$ are dual
basis of the orthogonal decomposition,
\be
\tilde{Z^b_{\mu}}~Z^\mu_a~=~\D^b_a,~~~~~
\tilde{Z^B_{\mu}}~Z^\mu_A~=~\D^B_A,~~~~~{\rm thus}~~~~~~
Z^\mu_a~\tilde{Z^a_{\nu}} ~+~ Z^\mu_A~\tilde{Z^A_{\nu}}=~
\D^\mu_\nu.
\label{orth}
\ee
$\tilde Z^{a}_\mu(\phi)$ 
and $Z^{\mu}_A(\phi)$'s are local operators
but $\tilde Z^{A}_\mu(\phi)$ 
and $Z^{\mu}_a(\phi)$'s are in general 
non-local since 
they verify the inhomogeneous differential equations 
\bref{orth}.
Thus the splitting of the 
parameters \bref{parameter} into the 
anomalous and non-anomalous ones 
is not performed locally in general.
Since  we should work in 
the space of local functionals we will not use
the explicit decomposition \bref{parameter};
it also happens that the covariance prevents 
the explicit decomposition,
as will be seen in the example of next section.
Here we should point out that we can still 
consider {\it in a local way 
the non-anomalous transformations} defined in \bref{system1}. 

\medskip

\subsection{Derivation of the WZ term}
\indent

The WZ term for a  gauged effective field theories of \NG 
bosons or for  
 anomalous gauge theories
 is a local functional satisfying  $\delta {\cal M}_1=i{\C{A}}$
with the anomaly satisfying the WZ consistency condition 
$\delta{\C{A}}=0$.
The anti-field independent part is found as a functional of 
the 
fields $\phi^i$ and \NG fields $\theta^\mu$ and verifies 
\be
\delta_\mu {\cal M}_1(\phi,\T^\nu)= i {\C{A}}_{\mu}(\phi).
\label{defwz1}
\ee
Remember that some of ${\C{A}}_{\mu}$ may be identically zero and some are 
dependent.
The solution of \bref{defwz1} is found in the extended space of 
$\phi^j$ 
and $\T^\nu$ as a sum of a particular solution 
and the general solution of 
the homogeneous equation. A particular solution, in a form of 
non-local 
functional ${\C{M}}_{1}^{non}(\phi)$, always exists since 
the BRST cohomology
is trivial in the space of non-local functionals. 
In physical terms there 
always exists a non-local counter term which cancels the 
anomaly.
It is an invariant function under \NATs of $\phi^j$. 
The solution of the 
homogeneous equation is obtained by making use of the 
finite gauge 
transformation $F^j(\phi^i,\T^\mu)$ of $\phi^j$ 
under ${\cal G}$. 
The \NG fields $\T^\mu$
are functions parametrizing the finite transformations. 
Some general properties of  quasigroup structures
are summarized in Appendix A.

The general solution of \bref{defwz1} is 
 \be
 \label{gensol}
 {\C{M}}_{1}(\phi,\T^\nu)={\C{M}}_{1}^{non}(\phi)+G(F^j(\phi,\T^\nu))
 \ee
if the transformation of $\T^\nu$ is determined in such 
a way that
\be
\delta_\ep F^j(\phi^i,\T^\mu)=0.
\label{cond1}
\ee
Since the  fields transform as 
$\D_\ep \phi^j=R^j_\nu(\phi)\ep^\nu$, 
the transformation of the NG fields is deduced, 
using \bref{eqtrB} of Appendix A, as 
\be
\D_\ep \T^\nu=~-{\tilde{\mu}_\s}^{\nu}(\T;\phi)\ep^\s,
\label{delTep}
\ee
where ${\tilde{\mu}_\s}^{\nu}(\T;\phi)$ is defined using the
composition function $\varphi^\nu(\T^\mu,{\T^{'\mu}};\phi^i)$ of
the two transformations in ${\cal G}$ as in \bref{vecZA}
\be
 \label{vecZA1}
  \tilde \mu^\nu_{~\s}(\T;\phi)~=~\left.\frac{\partial\vp^\nu(\T',\T;\phi)}
 {\partial\T'^\s}\right|_{\T'=0}.
 \ee

On the other hand the WZ term should verify the one-cocycle 
condition \cite{wz}
 \be
 \label{cocycle}
 {\C{M}}_{1}(\phi^j,\varphi^\nu(\T,{\T'};\phi))~=~{\C{M}}_{1}
 (F^j(\phi,\T),{\T^{'\nu}})~+~{\C{M}}_{1}(\phi^j,\T^\nu).
 \ee
It determines the arbitrary function $G(\phi)$ in  \bref{gensol}
and the solution satisfying the one-cocycle condition is 
\be
{\cal M}_{1}(\phi^j,\T^\nu)=\, {\cal M}^{non}_{1}(\phi^j)
- {\cal M}^{non}_{1}(F^j(\phi,\T)).
\label{WZ12}
\ee

Since we have introduced the NG fields for the non-anomalous 
transformations as well as for the anomalous ones the WZ 
term \bref{WZ12} 
is expected to have gauge invariances.
In fact by taking the variation of the WZ term 
with respect to $\T^\nu$, we have
\bea
\delta_\T{\cal M}_{1}(\phi,\T)=-\delta_\T {\cal M}^{non}(F)
=-i {\cal A}_{\nu}(F) \lambda^\nu_{~\s}(\T;\phi)\delta \T^\s
\label{motionwz}
\eea
where we have used the Lie equation \bref{eqtr} and 
$\lambda^\nu_{~\s}(\T;\phi)$ is an inverse of 
$\mu^\s_{~\nu}(\T;\phi)$ introduced in \bref{mu2};
\be
 \mu^\nu_{~\s}(\T,\phi)= \left.\frac{\partial\vp^\nu(\T,\T',\phi)}
{\partial\T'^\s}\right|_{ \T'=0} \, .
 \label{mu1}
\ee
The WZ term ${\cal M}_{1}(\phi^j,\T^\mu)$ is invariant under 
\be
\D_\s \T^\nu~=~{\xi}^{\nu}_A(\T;\phi)\sigma^A~. 
\label{delTsg}
\ee
where $\sigma^A$'s are the parameters of 
the gauge transformation 
and ${\xi}^{\nu}_A$ is defined using the local operator 
$Z^\r_A(\phi)$ in
\bref{NAtrans} by 
\be
{\xi}^{\nu}_A(\T;\phi)~\equiv~{\mu}^{\nu}_\r(\T;\phi) Z^\r_A(F(\T;\phi)).
\label{xi}
\ee 
Actually the $F^i(\phi,\theta)$ transforms under \bref{delTsg}
as a non-anomalous transformation, 
\be
\delta F^i(\phi,\theta)=R^i_\mu(F)Z^\mu_A(F)\sigma^A
\ee
and the WZ term \bref{WZ12} is invariant under 
arbitrary infinitesimal $\s^A$.

Combining \bref{delTep} and \bref{delTsg}, the transformation
property of the NG fields $\T^\nu$ is 
\be
\D ~\T^\nu=~-{\tilde{\mu}_\mu}^{\nu}(\T;\phi)\ep^\mu
+{\xi}^{\nu}_A(\T;\phi)\sigma^A.
\label{delTepf}
\ee
Note the transformation properties of the NG fields is 
in general non-linear 
due to the structure of $\tilde\mu(\theta,\phi)$ and $\mu(\theta,\phi)$. 
Furthermore as we will see 
below the 
algebra of these transformations closes only on shell of the 
equations of motion of the WZ term. 

Now let us rewrite the WZ term \bref{WZ12} as a surface 
integral in a variable $t$ 
\bea
\nonumber
{\cal M}_{1}(\phi^j,\T^\nu)&&= - \int_0^1 {d t} \, \frac{d}{d t} {\cal
M}^{ non}_{1}(F(\phi,\T_t))
\\
&& = - \int_0^1 {d t} \,\frac{\pa{\cal M}^{non}_{1}(F(\phi,\T_t))}{\pa F^i}
\frac{\pa F^i(\phi,\T_t)}{\pa \T^\nu_t}(\frac{\pa \T^\nu_t}{\pa t})
\eea
where $\T^\nu(t)\equiv \T^\nu_t$ is any interpolating function satisfying
the boundary conditions
\be
\T^\nu(1)~=~\T^\nu,~~~~~~\T^\nu(0)~=~\T^\nu_0={\rm identity~~transformation}.
\ee
Using the Lie equation \bref{eqtr} we obtain 
\be
{\cal M}_{1}(\phi^j,\T^\nu)
= - \int_0^1 {d t} \, \left.\{\frac{\pa{\cal M}^{non}_{1}(\phi)}
{\pa\phi^i}R^i_\nu(\phi)\}\right|_{\phi=F(\phi,\T_t)} 
\lambda^\nu_\r(\T_t;\phi) (\pa_t \T^\r_t).
\label{wzf1}
\ee
It can be expressed in terms of the anomaly as
 \be
 \label{WZ AI Cl}
 {\cal M}_{1}(\phi^j,\T^\nu) = \, -i \int_0^1 {d t} \,
 {\cal A}_{\nu}(F(\phi,\T_t)) \lambda^\nu_{~\r}(\T_t;\phi) (\pa_t \T^\r_t).
 \ee

Note that this expression is valid for a gauged effective theory of Goldstone
bosons associated with a broken global symmetry and for anomalous gauge 
theories.

\medskip

\subsection{The Extended Action}
\indent

In order to study the  gauge structure of the 
extended formalism of the fields $\phi^i$ and $\theta^\mu$ 
in more detail it is useful to introduce a field-antifield 
formalism 
and the classical master equation (CME) that  encodes all  
gauge structure of the theory \cite{bv2}\cite{fh}.
The  solution of CME is constructed in the extended phase space 
of
$(\phi^j,c^\nu,\T^\nu,b^A)$ and their anti-fields 
$(\phi^*_j,c^*_\nu,\T^*_\nu,b^*_A)$.  
The original solution $S$ of CME, $(S,S)=0$, is 
\be
S~=~S_0(\phi^j)~+~\phi^*_j~R^j_\nu(\phi)~c^\nu~+~\frac12
c^*_\nu~T^\nu_{\rho\s}(\phi)~c^\s c^\rho,
\label{cme}
\ee
where $T^\nu_{\rho\s}(\phi)$ is the structure function of $\CG$
depending on $\phi$ in case of quasigroups.
The extended action $\tilde S$ of the system is
\be
\tilde S~=~S~+~\T^*_\nu~\{-{\tilde{\mu}}^\nu_\s(\T;\phi)
c^\s~+~\xi^\nu_A(\T;\phi)b^A\}~+~
\frac12 b^*_A  \tilde T^A_{BD}(F(\phi,\T))b^D b^B,
\label{extaction}
\ee
where 
\be
{\tilde T}^{\tilde\r}_{{\tilde\mu}{\tilde\nu}}(\phi)~=~
\tilde Z^{\tilde\r}_\r~T^\r_{\mu\nu}~Z^\mu_{\tilde\mu}~
Z^\nu_{\tilde\nu}~-~\tilde Z^{\tilde\r}_\r~
\frac{\pa Z^\r_{[\tilde\nu}}{\pa \phi^i}~R^i_\s~Z^\s_{\tilde\mu]}.
\ee
The second term of \bref{extaction} comes from the transformation properties 
of the $\T^\nu$ in \bref{delTepf}. 
$b^A$ is ghost for the $\s^A$ gauge transformation of the WZ term.
The last term comes from the algebra of the 
$\s^A$ transformations in \bref{delTepf}.
The derivations are given in the Appendix B.
The $\tilde S$ satisfies 
\be
(\tilde S,\tilde S)~=~-\{\tilde \T^*_\mu~\mu^\mu_\nu~Z^\nu_a(F)~+~
b^*_A~\tilde T^A_{aB}(F)~
 b^B\}~\tilde T^a_{DE}(F)b^E  b^D.
\label{SS}
\ee
As will be shown the  $\tilde T^a_{DE}(F)$ vanishes using the WZ equation of 
motion the $\tilde S$ verifies the CME only on shell. The spinning
string with a super-\Diff invariant regularization shows an example of this
phenomena \cite{gkk}.

In order to see $\tilde T^a_{DE}(F)$ vanishes on shell\footnote{We acknowledge 
discussions with J.M. Pons and F. Zamora clarifying this point. } 
consider the fact that the effective action is invariant 
under \NATs, i.e.,
\be
\D\Gamma~=~\frac{\pa\Gamma}{\pa\phi^i}~R^i_\mu~(\phi)
Z^\mu_A(\phi)\tilde\ep^A~=~0
\ee
and therefore
\bea
0~=~[~\D(\ep), \D(\ep')~]~\Gamma~=~
\tilde{\cal A}_a(\phi)\tilde T^a_{AB}(\phi)\tilde\ep^A\tilde\ep'^B.~
\eea
Thus $\tilde T^a_{AB}(\phi)$ is a linear combination of 
the anomalies with anti-symmetric coefficients 
$E^{ab}_{AB}(\phi)=-E^{ba}_{AB}(\phi)$,
\be
\tilde T^a_{AB}(\phi)~=~E^{ab}_{AB}(\phi)\tilde{\cal A}_b(\phi).
\label{TaAB}
\ee
On the other hand the equations of motion for the \NG fields are seen
from \bref{motionwz}, using the fact that $\lambda$ is non-singular, as
\be
{\cal A}_\mu(F)~=~0.
\label{AF}
\ee
Thus $\tilde T^a_{AB}(F)$ vanishes on shell of WZ equation of motion.
\medskip

So far we have introduced the \NG fields $\T^\mu$ for all transformations.
It is not always necessary when there is a subgroup $\CG_0$ of $\CG$
and $\CG_0$ contains all \ATs. In this case the NG fields $\T^J$ 
associated with the transformation in $\CG-\CG_0$ and the corresponding 
ghosts $b^J$ can be eliminated consistently. 
The reduction is proceeded by using a canonical transformation in Appendix
C. The crucial point is that the BRST transformations for $\T^J$ is nilpotent 
on the reduced space defined by $\T^J=0$. 
As a result the transformation property of the remaining NG fields is 
modified from \bref{delTepf} to one in \bref{delTalpredA},
\be
\D \T^\A~=~\{-{\tilde{\mu'}}^{\A}_\nu~c^\nu~
+~\xi^\A_A~b^A\},~~~~~~~~~~
{\tilde{\mu'}}^{\A}_I~\equiv~{\tilde{\mu}}^{\A}_I-\mu^\A_J
\lam^J_L\tilde\mu^L_I,~~~
{\tilde{\mu}}^{\A'}_\B~\equiv~{\tilde{\mu}}^{\A}_\B.
\label{delTalpred}
\ee
where indices $\A,\B,...$ are those of $\CG_0$
and $I,J,...$ are non-anomalous indices of $\CG-\CG_0$.


\subsection{ Relation to $G/H$ non-linear realization
for Lie groups}
\indent

If the theory we are considering is such that the non-anomalous 
transformations close between themselves, $\tilde T^a_{AB}=0$,
 \bref{SS} tells $\tilde S$ satisfies the CME off shell thus 
the commutators of the $\s$ transformations of the NG fields 
is closed. 
If furthermore we can split in a local way the anomalous and \NATs
 we can eliminate the NG fields $\T^A$ associated to
the non-anomalous transformations explicitly. 
This is the case for YM Lie groups . As 
we will see our procedure gives a generalization of $G/H$ 
non-linear realization method of YM Lie groups to the case of quasigroups.

Let us restrict to the surface $\T^A=0$. In order to 
have consistent transformations on this surface
we should impose $\delta\theta^A|_{\theta^A=0}=0$.
Using \bref{delTepf} the parameters of $\sigma$ transformations are
determined in terms of the parameters $\epsilon^\mu$ 
\be
\sigma^B={\Lambda}^B_A(\T^a,\phi^j)~{\tilde\mu}^A_\nu(\T^a,\phi^j)~\epsilon^\nu
\label{detsig}
\ee
where ${\tilde\mu}^A_\nu(\T^a,\phi^j)\equiv{\tilde\mu}^A_\nu(\T^\nu,\phi^j)|_
{\theta^A=0}=0$ and ${\Lambda}^B_A$ is the inverse matrix of
$\xi^A_B|_{\theta^A=0}$.
From \bref{delTepf} and \bref{detsig} we deduce the 
transformation law for the NG fields $\T^a$ associated with the 
anomalous transformations as
\be
\D \T^a=~-[{\tilde{\mu}_\nu}^{a}(\T^a,\phi^j)~-~
{\xi}^a_A(\T^a,\phi^j) {\Lambda}^A_B(\T^a,\phi^j)
{\tilde\mu}^B_\nu~(\T^a,\phi^j)]~\epsilon^\nu.
\label{varcoset}
\ee 
This is a non-linear realization of the quasigroup in terms 
of the NG fields
$\T^a$ {\it and} $\phi^j$. 
In this case the WZ term is given as a function of $\phi$ and $\T^a$,
by putting $\T^A=0$ as
 \be
 {\cal M}_{1}(\phi,\T^a) = \, -i \int_0^1 {d t} \,
 {\cal A}_\nu(F(\phi,\T^a_t)) \lambda^\nu_{~b}(\T^a_t;\phi) 
(\pa_t \T^b_t).
 \ee
There is no gauge symmetry in  this WZ term.

In the reduced variables the one-cocycle condition corresponding to 
\bref{cocycle} holds also as
 \be
 \label{cocyclered}
 {\C{M}}_{1}(\phi,\tilde\varphi^d(\T^a,{\T^{'b}};\phi))~=~{\C{M}}_{1}
 (F(\phi,\T^a),{\T^{'b}})~+~{\C{M}}_{1}(\phi,\T^a).
 \ee
Here the composition of two anomalous transformations 
$\varphi(\T^a,{\T^{'b}};\phi)$ produces 
non anomalous components $\varphi^A$
as well as anomalous one $\varphi^d$. However it is expressed as a 
composition of an anomalous transformation $\tilde\varphi^d$ and 
successive non-anomalous one and since the WZ term is invariant under 
the \NATs we obtain the result \bref{cocyclered}. 
\medskip

In the case of a YM Lie group $G$ with a reductive non-anomalous subgroup 
$H$, the formula \bref{varcoset} does not depend on the  fields 
$\phi$ and reproduces the infinitesimal transformation of the NG fields 
$\T^a$ of the coset ${G}/{H}$ obtained from the standard method 
of non-linear realizations \cite{w}\cite{cwz}.
In this case $\tilde\mu$ and $\mu$ respectively 
give the left and the right (infinitesimal) transformations of the group 
element. The group element $g$ is parametrized using some representation 
$t_\mu$ of $\CG$ as $g=e^{\T^\mu t_\mu}$.
If we parametrize left transformations as $g_L=e^{-\ep^\mu t_\mu}$ 
and right transformations as $g_R=e^{\s^\mu t_\mu}$ then
$~g$ transforms to $ ~g'=g_L~g~g_R~$. 
If we define the composition function by 
$gg'=e^{\T^\mu t_\mu}e^{{\T'}^\mu t_\mu}=e^{\vp(\T,\T')^\mu t_\mu}$
the infinitesimal transformation of $\T^\nu$ is given by
\be
\D\T^\r~=~-\tilde\mu^\r_\nu(\T)\ep^\nu~+~\mu^\r_\nu(\T)\s^\nu,
\ee
where $\mu$ and $\tilde\mu$ are defined as \bref{mu1} and \bref{vecZA1}
though they are independent of $\phi$.

The non-linear realization is defined by restricting $\T^\r$ to be the coset
coordinates $\T^a$, i.e. the NG fields, and are transformed under the left 
action of $G$ with the associative right action of $H$. The latter is 
determined so that the $\T'^\r$ remains in the coset. 
For infinitesimal transformations the condition is 
\be
0~=~\D\T^A~=~\left.\{-\tilde\mu^A_\nu(\T)\ep^\nu~+~\mu^A_B(\T)\s^B\}\right|_
{\T^A=0},~~~~~\rightarrow~~~~~
\sigma^B={\Lambda}^B_A~{\tilde\mu}^A_\nu~\epsilon^\nu
\label{detsigap}
\ee
where ${\Lambda}^B_A$ is the inverse matrix of
$\mu^A_B|_{\theta^A=0}$. It follows the transformation law for the 
NG fields $\T^a$ 
\be
\D_\ep \T^a=~-[{\tilde{\mu}_\nu}^{~a}(\T^a)~-~
{\mu}^a_A(\T^a) {\Lambda}^A_B(\T^a)
{\tilde\mu}^B_\nu~(\T^a)]~\epsilon^\nu
\label{varcosetap}
\ee 
which is one corresponding to our general quasigroup result
\bref{varcoset}. Note in this case we always split the \ATs and \NATs
thus the  $Z$ and $\tilde Z$ in \bref{orth} are taken to be unit matrices and
$\xi^\r_A=\mu^\r_A$. In the case the anomalous transformations from a
subgroup the matrix ${\Lambda}^A_B(\T^a)$ coincides with 
$\left.{\lambda}^A_B(\T^\nu)\right|_{\T^A=0}.$

Using the explicit forms of $\mu$ and $\tilde \mu$ in the ajoint representation
\be
\mu^\r_\s=(\frac{\hat\T}{e^{\hat\T}-1})^\r_\s,~~~~
\tilde\mu^\r_\s=(\frac{-\hat\T}{e^{-\hat\T}-1})^\r_\s,~~~~
{\hat\T}^\r_\s\equiv \T^\nu T^\r_{\nu\s}
\ee
we can see how the transformations associated with $H$ produces linear 
transformations for the NG fields while 
the ones associated with the coset are non-linear.

\section{Weyl Invariant 2D Gravity}
\indent

In this section we apply the general formula developed in the 
previous section to the
case of 
 gravity in two dimensions. We use zwei-bein formalism for the gravitational
fields.
The classical
gauge symmetries are thus the local Lorentz, the Weyl and the two 
dimensional \Diff (Diff$_2$) transformations.
The two dimensional gravity coupling to scalar matter (bosonic string) 
has been discussed as a system with anomalous Weyl symmetry \cite{p}.
This result comes from the regularization preserving the diffeomorphism
invariance. Recently a Weyl invariant formulation has been 
discussed\cite{kmk}\cite{j}\cite{abs}.
The anomaly is
\be
{\cal A}~=~k \int d^2x~ R(\frac{g_{\mu\nu}}{\sqrt{-g}})\pa_\A c^\A~=~
{\cal A}_\A~ c^\A,
\label{diffano}
\ee
where $R(\frac{g_{\mu\nu}}{\sqrt{-g}})$  is a scalar curvature of the
Weyl invariant metric.
The \Diff ghosts $c^\A,(\A=0,1)$ appear in a form of divergence and 
the components of anomaly ${\cal A}_\A$ are not independent, we see that
 the area preserving \Diff (SDiff) is the \NAT. 
The corresponding infinitesimal transformation parameters are satisfying  
$\pa_\A\ep^\A=0$, they form a sub-algebra of the \Diff transformations.
Ones that do not satisfy $\pa_\A\ep^\A=0$ generate \ATs. 

The splitting of the \Diff transformations into the anomalous and \NATs
 cannot be performed locally and covariantly in this case.
In constructing the WZ term we take \Diff as the ${\cal G}_0$,
a subgroup including all anomalous transformations,
we do not need to introduce the NG fields for the local Lorentz and the Weyl 
transformations. For Diff$_2$ we consider finite coordinate transformations
\be
x^\A~\rightarrow~\tilde x^\A~=~f^\A(x)
\ee
and the functions $f^\A(x),~(\A=0,1)$ play the role of the NG fields
$\T^\A$. The corresponding finite transformation of zwei-bein is
\be
e_\A^{~a}~\rightarrow~F_\A^{~a}(e,f) ~=~
\frac{\pa f^\B(x)}{\pa x^\A}e_\B^{~~a}(f(x))
\equiv~A_\A^{~~\B}(x)~e_\B^{~~a}(f(x)).
\label{tilde}
\ee 
The determinant $e=\det e_\A^a$ is transformed as 
\be
e(x)~\rightarrow~\tilde e(x)~=~\Delta^f(x)e(f(x)),~~~~~
\Delta^f(x)\equiv\det A=\det~(\pa_\A f^\B(x)).
\ee 

The successive two finite transformations gives the composition 
function $\varphi(f,f')$ as
\be
\varphi^\A(f,f')~=~f''^\A(x)~=~f^\A(f'(x)).
\label{compf}
\ee
 
$\tilde\mu^\A_{\B}(f)$ in \bref{vecZA1} has a non-local expression,
\be
\tilde\mu^\A_{\B}(f)~=~\left.\frac{\pa \varphi^\A(f',f)(x)}{\pa f'^\B(x')}
\right|_
{f'=x'}~=~\delta^2(x'-f(x))\D^\A_{\B}.
\label{mutnon}
\ee
On the other hand the functions $\mu$ in \bref{mu1} and its inverse $\lam$ 
have local forms, 
\be
 \mu^\A_\B(f)= \left.\frac{\partial
\vp^\A(f,f')}{\partial f'^\B(x')}
 \right|_{ f'=x'} ~=~ {\pa_\B f^\A(x)}\delta(x'-x),
\ee
\be
{\lambda_{\A}^{\B}}(f) ~=~{{A_\A^{-1\B}}}(x)\delta(x'-x)~=~
\frac{1}{\Delta^f}\ep^{\B\G}(\pa_\G f^\D)\ep_{\D\A}\delta(x'-x).
\label{lambda2}
\ee

In order to find the expression for the WZ term we consider an interpolating
variable $f_t^\A(x)$ satisfying the boundary conditions
\be
f_0^\A(x)~=~x^\A ~~~(t=0),~~~~~
f_1^\A(x)~=~f^\A(x) ~~~(t=1).~~~~~
\label{boundaryf}
\ee
The finite transformation of the curvature in the anomaly 
\bref{diffano}, which is evaluated by a Weyl invariant combination of 
zwei-bein $\frac{e_\mu^{~a}}{\sqrt{e}}$, is
\be
R(\frac{{e^{f_t}}_\mu^{~a}(x)}
{\sqrt{{e^{f_t}}(x)}})~=~
\Delta^{f_t}(x)[\left.\{~R(\frac{e_\mu^{~~a}}{\sqrt{{e}}})~+~
\Box \ln{\Delta^{f_t}(F_t(x))}\}\right|_{x=f_t(x)}],
\label{Rft}
\ee
where $F_t^\A(\hat x)$ is an inverse function of $f_t^\A(x)=\hat x^\A,~i.e.~
x^\A=F_t^\A(f_t(x))$. 
$\Box$ is defined as $\Box\equiv\pa_\A(eg^{\A\B}\pa_\B)$.

Now we are ready to calculate the WZ term \bref{WZ AI Cl}
\bea
 {\cal M}_{1}(\phi,f) = \, -i \int_0^1 {d t} \,
 {\cal A}_{\A}(F(\phi,f_t)) \lambda^\A_{~\B}(f_t,\phi) 
(\pa_t f_t^\B)~~~~~~~~~~~~~~~~~~~~~~~~~~
 \nn\\
= \, -i \int_0^1 {d t} \,\int d^2 x~\Delta^{f_t}[~R(\frac{{e}_\mu^{~a}}
{\sqrt{{e}}})~+~\Box\ln{\Delta^{f_t}(F_t(x))}]_{x=f_t(x)}~
\pa_\A~[{{A_{\B t}^{-1\A}}}(x)~{\dot f}_t^\B(x)],
\label{WZ2}
\eea
where ${\lambda_{\G}^{\B}}$ is given in \bref{lambda2}. 
We change the integration variable from $x^\A$ to $\hat x^\A=f_t^\A(x)$
,it becomes
\bea
{\cal M}_{1}(\phi,f) = \, -i \int_0^1 {d t} \,\int d^2 \hat x~
[~R(\frac{{e}_\mu^{~a}}{\sqrt{{e}}})~+~\Box\ln{\Delta^{f_t}(F_t)}
]_{x=\hat x}
\nn\eea
\be~
\times[\pa_\A~\{{{{A_{\B t}}^{-1\A}}}(x)~{\dot f}_t^\B(x)\}]_
{x=F_t(\hat x)}.
\ee
The last factor in the integrand is expressed as a total derivative 
with respect to $t$
\be
[\pa_\B\{A_{\A t}^{-1\B} \dot f_t^{\A}\}]_{x=F_t(\hat x)}~=~
\pa_t[\ln(\Delta^{f_t})_{x=F_t(\hat x)}].
\ee
Defining the expression appearing in the r.h.s. as $\Th_t(\hat x)$;
\be
\Th_t(\hat x)=\ln\left.\Delta^{f_t}(x)\right|_{x=F_t(\hat x)}
\ee
it becomes
\be
{\cal M}_{1}(\phi,f)~= \, -i \int_0^1 {d t} \,\int d^2 \hat x~
[~R(\frac{{e}_\mu^{~a}(\hat x)}
{\sqrt{{e}(\hat x)}})~+~\Box_{\hat x}\Th_t(\hat x)]
\pa_t\Th_t(\hat x).
\ee
The $t$ integration in the WZ term is performed and we obtain 
\be
{\cal M}_{1}(\phi,f)~=~-i\int d^2 x~[~R(\frac{{e}_\mu^{~a}}
{\sqrt{{e}}})~\Th~-~\frac12\{e~g^{\A\B}~\pa_\B~\Th~\pa_\A~\Th\}],
\label{WIWZ}
\ee
where we have used the boundary condition \bref{boundaryf}; 
$\Th_{t=0}(\hat x)=0$ ~and we call 
$\Th_{t=1}(\hat x)$ as $\Th(\hat x)$;
\be
\Th(\hat x)~\equiv~\ln\left.\Delta^{f}(x)\right|_{x=F(\hat x)}.
\label{Theta}
\ee
The WZ term \bref{WIWZ} is equivalent to one found in \cite{kmk}.
\vskip 6mm

We will make several comments.
The variable $\Th(x)$ is a non-local function of $f^\A(x)$.
The WZ term, which is non-local as a function of $f^\A(x)$,
can be expressed only in term of one variable $\Th(x)$ 
and becomes a local function of $\Th(x)$. 

The transformation properties of $f^\A(x)$ is determined by \bref{delTalpred}.
Now 
\be 
\tilde{\mu'}^\A_I~=~0,~~~
\xi^\A_A~=~\mu^\A_\nu Z^\nu_A~=~\pa_\B f^\A(x)\ep^{\B\G}\pa_\G.
\ee
Since $\tilde{\mu}^\A_\B$ given in \bref{mutnon} is 
non-local, $f^\A(x)$ transforms non-locally as
\be
\D f^\A(x)~=~-{\tilde{\mu'}}^\A_\mu(f)~\ep^\mu~+~
\xi^\A_A(f)\s^A~=~ 
-\ep^\A(f(x))+\pa_\B f^\A(x)\ep^{\B\G}\pa_\G \s(x),
\label{deltaf}
\ee
where $\s$ is a parameter of the additional gauge symmetry of the WZ action. 
The transformation of $\Th$ is given using that of $f^\A$ \bref{deltaf} and 
has a local form
\be
\D \Th(x)~=~\pa_\A\Th(x)\ep^\A(x)-\pa_\A\ep^\A(x).
\label{delTh}
\ee
That is it transforms as a scalar under SDiff satisfying $\pa_\A\ep^\A=0$. 
The WZ term \bref{WIWZ} in terms of $\Th$ is manifestly gauge invariant
under $\s-$ transformations due to the invariance of the $\Th$. 

Note that the 
inhomogeneous transformation of  $\Th$ suggests, according to the non-linear
realization procedure, that $\Th$ is an element of the coset 
$\frac{Diff_2}{SDiff}$. In fact, 
suppose $f_1^\A(x)$ and  $f_2^\A(x)$ belong to the equivalent class.
That is they are connected by a SDiff transformation by $f^\A(x)$;
\be
f_1^\A(x)~=~f_2^\A(f(x))~~~~~~{\rm with}~~~~~~~\Delta^f(x)~=~1.
\ee
Their determinants are related by $\Delta^{f_1}(x)=\Delta^{f_2}(f(x))$.
The definition of $\Th$, \bref{Theta}, enables us to  show that $\Th$ is a 
representative
of the coset
\bea
\Th_2(\hat x)&=~\left.\ln\Delta^{f_2}(x)\right|_{x=F_2(\hat x)}~=~
  \left.\ln\Delta^{f_1}(x)\right|_{f(x)=F_2(\hat x)}
\nn\\
&=~\left.\ln\Delta^{f_1}(x)\right|_{x=F(F_2(\hat x))}~=~ 
   \left.\ln\Delta^{f_1}(x)\right|_{x=F_1(\hat x)}~=~ \Th_1(\hat x). 
\eea
That is $\Th_1(\hat x)$ and $\Th_2(\hat x)$ have the same value at identical
point $\hat x$. 

Note also that by
 construction the WZ term satisfies the cocycle condition. It is manifest 
if it is expressed in terms of finite function $f^\A$. However it is not 
apparent once it is expressed in terms of $\Th$ as \bref{WIWZ}. 
The one-cocycle condition is now expressed as
 \be
 \label{cocycleTh}
 {\C{M}}_{1}(\phi,\varphi(\Th,\Th'))~=~{\C{M}}_{1}
 (F(\phi,\Th),\Th')~+~{\C{M}}_{1}(\phi,\Th).
 \ee
Since we have the relation between $\Th$ and  $f^\A(x)$ in \bref{Theta},
we can find the composition function of $\Th$'s
using the composition rule \bref{compf}, $f''(x)=f(f'(x))$,
\be
\Th''(\hat x)~\equiv~\varphi(\Th,\Th')~=~
\Th(\hat x)~+~[\Th'(x')]_{x'=F(\hat x)}.
\ee
The composition law has non-local form since $\Th'$ is evaluated at
${x'=F(\hat x)}\not=\hat x$, from which we can check the cocycle
condition \bref{cocycleTh}  explicitly.

\vskip 6mm  

The extended action $\tilde S$ in \bref{extaction} of this system is
\be
\tilde S~=~S~+~
\int dx [~f^*_\A\{-C^\A(f(x))+\pa_\B f^\A\ep^{\B\G}\pa_\G b~\}~+~
\frac12 b^*\ep^{\A\B}\pa_\A b\pa_\B b~],
\ee
where the first term  $S$ is the solution of the CME \bref{cme}
given by 
\bea
S~&&=~S_0~+~\int dx \{X^*~\partial_\A X~C^\A~ +~
{e^\A_{~a}}^*~(\partial_\B e_\A{}^a~C^\B
+\ep^a{}_b~e_\A{}^b~C_L+e_\A{}^a~C_W)
\nn \\
&& +~{C_\A}^*~\partial_\B C^\A~C^\B~ +~ {C_L}^*~\pa_\A C_L~C^\A
 +~{C_W}^*~\partial_\A C_W~C^\A ~\},
\eea
and $b$ is the ghost for the additional symmetry of the WZ action;
SDiff. 
It is non-local in terms of $f^\A(x)$.  
Since the non-anomalous transformations are closed in this case
the extended action $\tilde S$ satisfies the CME; $(\tilde S,\tilde S)~=~0$.
\medskip

The anomalous degree of freedom of $f^\A$ has been expressed as $\Th$. 
We can show 
that the other (non-anomalous) degree of freedom, say $\Th'$, 
and the additional
ghost $b(x)$ are transformed into a form of trivial non-minimal pairs. 
It is performed by making a canonical transformation.
$\Th'$ can be any function $w(f)$ as long as it is independent of $\Th(f)$.
The generating function is 
\bea
W&&=~\int dx~[~\Th^*(x)\ln(\Delta^f)_{x=F(x)}~+~{\Th'}^*(x)w(f(x))
\nn\\
&&+~{\tilde b}^*(x)~\frac{\pa w(f)}{\pa f^\A}~\{-C^\A(f(x))~+~
\pa_\B f^\A\ep^{\B\G} \pa_\G b(x)~\}~+~ {\tilde C}^*_\A C^\A ~],
\eea
It defines new fields,
\bea
&&\Th(x)~=~\ln(\Delta^f)_{x=F(x)},~~~~~~
\Th'(x)~=~w( f(x)),
\nn\\
&&{\tilde b}(x)~=~w_\A(f)~\{-C^\A(f(x))~+~\pa_\B f^\A \ep^{\B\G}\pa_\G b(x)~\}
,~~~
{\tilde C}^\A(x)~=~C^\A(x)
\eea
and the corresponding anti-fields.
In terms of new variables, all ${\tilde b}^*$ terms cancel out and  we obtain
\be
\tilde S~=~\hat S~+~\int dx~[~\Th^*\{\pa_\A\Th{\tilde C}^\A-
\pa_\A{\tilde C}^\A~\}~+~{\Th'}^*{\tilde b}~],
\ee
where $\hat S$ is $S$ in which all fields and anti-fields are replaced by 
corresponding new ones. The second term tells the transformation of $\Th$
\bref{delTh}.
The third term shows that $({\Th'},{\Th'}^*)$ and 
$(\tilde b,{\tilde b}^*)$
are non-minimal pairs and are cohomologically trivial. 
Note the extended action became local one as the result of the non-local 
canonical transformation. 
\section{Summary}
\indent

In this paper we have discussed in a unified way the WZ terms 
for:
 gauged effective field theories 
associated with a  spontaneously broken global symmetry,
and  anomalous gauge theories. The formalism  is applicable 
for the quasigroup case
in which each the anomalous and \NATs do not form closed subgroups. 
In general we should introduce 
the \NG fields associated with the \NATs. 
In this case the WZ terms have (hidden) gauge symmetries in such a way 
to kill the extra \NG fields. 
Our procedure can be seen as a 
generalization of the non-linear realization method of Lie groups.

As an example,  we have considered  two dimensional gravity 
with anomalous \Diff, and obtained  
 the local WZ term  described in terms of the coset coordinate
of $\frac{Diff_2}{SDiff}$.

\vskip 6mm
{\bf Acknowledgments}

We thank J.M.Pons and F.Zamora for their collaboration in the early stages
of this work. J.G acknowledge helpful conversations with S. Weinberg
about Goldstone bosons and WZ terms.
 This work has been partially supported by
 Robert A. Welch foundation, NSFT PHY9511632, CYCYT under contract number 
AEN93-0695, Comissionat per Universitats i Recerca de la Generalitat 
de Catalunya and  Commission of European Communities contract 
CHRX-CT93-0362(04).
One of the author (KK) would thank "{\it Nukada Grant}" for financial
support. 

\vskip 6mm

\appendix
\section{Quasigroup Structures}
\indent

Here we give a brief summary of quasigroup structures and 
some definitions \cite{batalin}.
 Let us consider the action of quasigroup $\CG$ which is locally
 described by a set of  parameters $\theta^\mu$
 on a manifold ${\cal M}$ parametrized by the classical fields ${\phi^i}$
 \bea
 \nonumber
 & F: {\cal M} \times {\cal G}\rightarrow {\cal M}
 \\
 \label{actionF}
 &\quad\quad\quad\quad(\phi^i,~\theta^\mu) \mapsto F^i (\phi,\theta).
 \eea
We assume it is closed and irreducible for simplicity. 
The relevant properties are invariance of the action;
\be
 {\cal S}_{0}( F (\phi,\theta))={\cal S}_{0}(\phi)
\ee
and the composition law;
 \be
 F^i (F (\phi,\theta),\theta')=
 F^i (\phi,\varphi(\theta,\theta';\phi)),
 \label{openF}
  \ee
 where $\varphi^\mu(\theta,\theta';\phi)$ represents the composition
 function of the  parameters of quasigroup $\CG$.
\vskip 3mm

The structure functions appearing in the solution of the classical master
equation \bref{cme} are directly related to these functions,
\be
 {R^i}_{\mu}( \phi)~:=~ \left.\frac{\partial F^i(\phi,
 \T)}{\partial\T^\mu}\right|_{\T= 0},~~~~~
 T^{\r}_{\mu \nu}(\phi) :=-
 \left(\frac{\partial^2
 \varphi^{\r}(\theta,\theta';\phi)}{\partial\theta^\mu
 \partial\theta'^\nu} - (\nu \leftrightarrow \mu)
 \right)_{\theta=\theta'=0}.
\ee
They satisfy
 \be
 \label{RR}
 {R^j}_{~[\mu}(\phi) {R^i}_{\nu],j} (\phi) + T^{\r}_{\mu \nu}(\phi)
 {R^i}_\r (\phi)~=~0,
 \ee
\be
\label{4-4}
\sum_{P\in{\rm Perm}[\mu\nu\r]} (-1)^P(~T^\delta_{\nu\r,j}(\phi)~R^j_{~\mu}
(\phi)~+
T^\delta_{\s\r}(\phi)~T^\s_{\mu\nu}(\phi)~)~=~0.
\ee
The latter is the generalized Jacobi identity.
\medskip 

 From the composition law \bref{openF} we can obtain an analogue of the
 Lie equation by multiplying an operator
 $\left. \frac{\partial}{\partial{\T'}}
 \right|_{\T'=0}$ on it, 
\be
 \frac{\partial F^i(\phi,\T)}{\partial\T^\s} = R^i_\nu(F(\phi,\T))
 \lambda^\nu_{~\s}(\T,\phi),
 \label{eqtr}
 \ee
 where $\lambda^\nu_{~\s}(\T,\phi)$ is the inverse matrix of
 \be
 \mu^\s_{~\nu}(\T,\phi)= \left.\frac{\partial
\vp^\s(\T,\T',\phi)}{\partial\T'^\nu}
 \right|_{ \T'=0} \, .
 \label{mu2}
 \ee
 On the other hand if we operate  $\left. \frac{\partial}{\partial{\T}}
 \right|_{\T=0}$ on \bref{openF} we have
\be
 \frac{\partial F^i(\phi,\T)}{\partial \phi^k}~R^k_\s(\phi)~=~
 \frac{\partial F^i(\phi,\T)}{\partial \T^\nu}~\tilde \mu^\nu_{~\s}
(\T;\phi),
 \label{eqtrB}
 \ee
 where
  \be
 \label{vecZA}
  \tilde \mu^\nu_{~\s}(\T,\phi) :=\left.\frac{\partial\vp^\nu(\T',\T;\phi)}
 {\partial\T'^\s}\right|_{\T'=0}.
 \ee
\vskip 3mm

Some useful formulas are 
\bea
\tilde\mu^\r_\mu~ D_\r~\tilde\mu^\s_\nu-
\tilde\mu^\r_\nu~ D_\r~\tilde\mu^\s_\mu&=&\tilde\mu^\s_\r~ T^\r_{\mu\nu}(\phi),
\label{mm1}
\\
\mu^\r_\mu ~\pa_\r~\mu^\s_\nu-
\mu^\r_\nu ~\pa_\r~\mu^\s_\mu&=&-\mu^\s_\r ~T^\r_{\mu\nu}(F(\phi,\T)),
\label{mm2}
\\
\tilde\mu^\r_\mu ~D_\r~\mu^\s_\nu-
      \mu^\r_\nu ~\pa_\r~\tilde\mu^\s_\mu&=&0,
\label{mm3}
\eea
where 
$\tilde\lam^\s_\r$ is the inverse of $\tilde\mu^\r_\s$  and
$D_\r~\equiv~\frac{\pa}{\pa \T^\r}-R^j_\s(\phi)~\tilde\lambda^\s_\r~
\frac{\pa}{\pa \phi^j}~\equiv~
\pa_\r-R^j_\s(\phi)~\tilde\lambda^\s_\r~\pa_j$.
\medskip

\section{The Extended Action}
\indent

In order to study the  gauge structure of the extended formalism 
of fields $\phi$ and $\theta$
in more detail it is useful to introduce a field antifield formalism 
and the classical master equation (CME) that  encodes the 
gauge structure of the theory \cite{bv2}\cite{fh}.
The  solution of CME in the original space of
$(\phi^j,c^\mu)$ and their anti-fields $(\phi^*_j,c^*_\mu)$ is  
\be
S~=~S_0(\phi^j)~+~\phi^*_j~R^j_\mu(\phi)~c^\mu~+~\frac12
c^*_\mu~T^\mu_{\rho\s}(\phi)~c^\s c^\rho
\ee
and satisfies, $(S,S)=0$. Here we use  \bref{RR} and 
\bref{4-4}.

We first construct an action in the extended phase space, by
introducing ghosts $b^\mu$,

\be
S_1~=~S~+~\T^*_\mu~\{-{\tilde{\mu}}^\mu_\nu(\T;\phi)~
c^\nu~+~\mu^\mu_\nu(\T;\phi)b^\nu\}~+~
\frac12 b^*_\r  T^\r_{\mu\nu}(F(\phi,\T))b^\nu b^\mu,
\label{extaction1}
\ee
which satisfies CME $(S_1,S_1)=0$ using \bref{mm1}\bref{mm2}\bref{mm3}.

However the WZ term is not invariant under all $b^\mu$
transformations but 
\bea
\delta_\s{\cal M}_{1}&=-\delta_\s {\cal M}^{non}(F)=
-{\cal A}_\r(F)\lam^\r_\mu{\mu^\mu_\nu b^\nu}=
-\tilde{\cal A}_a(F) \tilde Z^a_\nu(F) b^\nu.
\label{motionwz1}
\eea
Thus the WZ term is invariant if $b^\nu$ take the form
\be
b^\nu~=~Z^\nu_A(F)~\tilde b^A.
\ee
Therefore we make a canonical transformation by
\be
W~=~b^*_\mu~Z^\mu_{\tilde \nu}(F)~\tilde b^{\tilde\nu}~+~
\phi^*_j~\tilde\phi^j~+~\T^*_\mu~\tilde\T^\mu~+~c^*_\mu~\tilde c^\mu,
\ee
where tilded indices, $\tilde\nu$, runs through non anomalous
ones, $A$, and anomalous ones, $a$. It gives
\be
b^\mu~=~Z^\mu_{\tilde \nu}(F)~\tilde b^{\tilde\nu}~\equiv~
Z^\mu_{A}(F)~\tilde b^{A}~+~Z^\mu_{a}(F)~\tilde b^{a},~~~~~
\tilde b^*_{\tilde \nu}~=~~b^*_\mu~Z^\mu_{\tilde \nu}(F)
\ee
\be
\tilde\phi^*_j~=~\phi^*_j~+~b^*_\mu \tilde b^{\tilde\mu}
\frac{\pa Z^\mu_{\tilde\mu}(F)}{\pa F^i}
\frac{\pa F^i}{\pa\phi^j},~~~~~
\tilde\T^*_\r~=~\T^*_\r~+~b^*_\mu \tilde b^{\tilde\mu}
\frac{\pa Z^\mu_{\tilde\mu}(F)}{\pa F^i}
\frac{\pa F^i}{\pa\T^\r}.
\ee
The action $S_1$ becomes
\be
S_1~=~\hat S~+~\tilde\T^*_\mu~\{-{\tilde{\mu}}^\mu_\nu(\T;\phi)~
c^\nu~+~\mu^\mu_\nu(\T;\phi)Z^\nu_{\tilde\nu}(F)\tilde b^{\tilde \nu}\}~+~
\frac12 \tilde b^*_{\tilde\r} {\tilde T}^{\tilde\r}_{{\tilde\mu}
{\tilde\nu}}(F(\phi,\T)){\tilde b}^{\tilde\nu} {\tilde b}^{\tilde\mu},
\label{extaction2},
\ee
where $\hat S$ is $S$ in which the old variables are replaced by the 
corresponding the new ones and 
\be
{\tilde T}^{\tilde\r}_{{\tilde\mu}{\tilde\nu}}(\phi)~=~
\tilde Z^{\tilde\r}_\r~T^\r_{\mu\nu}~Z^\mu_{\tilde\mu}~
Z^\nu_{\tilde\nu}~-~\tilde Z^{\tilde\r}_\r~
\frac{\pa Z^\r_{[\tilde\nu}}{\pa \phi^i}~R^i_\s~Z^\s_{\tilde\mu]}.
\ee
It satisfies the generalized Jacobi identity corresponding to \bref{4-4}.
\be
\label{tildejacobi}
\sum_{P\in{\rm Perm}[\mu\nu\r]} (-1)^P
(~\tilde T^\delta_{\nu\r,j}~R^j_{~\s}~Z^\s_\mu~+
\tilde T^\delta_{\s\r}~\tilde T^\s_{\mu\nu}~)~=~0.
\label{gji}\ee
Since we made a canonical transformation the $S_1$ satisfies the
CME also.

We are interested in an extended formalism which incorporates the
gauge invariances of the WZ term we must 
 impose $\tilde b^a=0$. It is consistent only on the 
WZ equation since
\be
\left.\D\tilde b^a~\right|_{\tilde b^a=0}~=~\frac12 \tilde T^a_{AB}(F)
\tilde b^B \tilde b^A
\ee
and $\tilde T^a_{AB}(F)$ vanishes on shell of WZ equation of motion,
see \bref{TaAB} and \bref{AF}.
Correspondingly the action 
\be
\tilde S~=~\left. S_1\right|_{\tilde b^a=0}~=~
\hat S~+~\tilde\T^*_\mu~\{-{\tilde{\mu}}^\mu_\nu(\T;\phi)~
c^\nu~+~\mu^\mu_\nu(\T;\phi)Z^\nu_{A}(F)\tilde b^{A}\}~+~
\frac12 \tilde b^*_{A} {\tilde T}^{A}_{{B}{D}}(F(\phi,\T))
{\tilde b}^{D} {\tilde b}^{B},
\label{extaction3}
\ee
satisfies CME on shell of the WZ equation.
\be
(\tilde S,\tilde S)~=~-\{\tilde \T^*_\mu~\mu^\mu_\nu~Z^\nu_a(F)~+~
\tilde b^*_A~\tilde T^A_{aB}(F)~
\tilde b^B\}~
\tilde T^a_{DE}(F)\tilde b^E \tilde b^D,
\label{nil}
\ee
where we have used \bref{gji}.

Rewriting new variables by omitting {\it tildes} ,e.g. 
$\tilde b^A~\rightarrow~b^A$, we find the extended action
\bref{extaction} and the on shell nilpotency \bref{SS} in the section 2.2.
\medskip


\section{Reduction of variables in $\CG-\CG_0$}
{\indent}

When there is a subgroup $\CG_0$ whose algebra includes all anomalous 
generators we can reduce the extra variables $\T^I$'s of $\CG-\CG_0$.
It is corresponding to define the Dirac brackets
in the canonical constrained systems. 

First we introduce the extra variable $\T^\mu$ for all transformations in 
$\CG$. The extended action $\tilde S$ of the system is \bref{extaction};
\be
\tilde S~=~S~+~\T^*_\mu~\{-{\tilde{\mu}}^\mu_\nu(\T,\phi)~c^\nu~+~
\xi^\mu_\tA(\T,\phi)b^\tA\}~+~\frac12 b^*_\tA \tilde T^\tA_{\tB\tD}(F)
b^\tD b^\tB,
\label{extaction0}
\ee
where indices $\tA,\tB,...$ are those of \NATs; $A,B,...$ of $\CG$ 
and $I,J,K,...$ of $\CG-\CG_0$.

To reduce the trivial non-anomalous variables $\T^I$ of $\CG-\CG_0$ we impose
\be
\T^I~=~0
\ee
and require
\be
0~=~\left.\D\T^I\right|_{\T^I~=~0}~=~
\left.\{-{\tilde{\mu}}^I_\nu~c^\nu~+~\xi^I_A~b^A+~\xi^I_J~b^J\}
\right|_{\T^I~=~0}.
\label{delth}
\ee
Since $\CG_0$ is the subgroup it holds, 
for indices $\A$ (A and a) of $\CG_0$ and for ${\T^I~=~0}$ 
\be
\mu^I_\A~=~\tilde\mu^I_\A~=~\lam^I_\A~=~\tilde\lam^I_\A~=~0.
\label{ancond}
\ee
Also $\T^I$ transformations of $\CG-\CG_0$ are \NATs decoupling from 
$\T^\A$ of  $\CG_0$ we can take $Z^\nu_I=\D^\nu_I$ and $Z^J_\A=0$
thus $\xi^I_A=\mu^I_\nu Z^\nu_A=0$ and $\xi^I_J=\mu^I_\nu Z^\nu_J=\mu^I_J$.
The condition \bref{delth} determines $b^I$;
\be
\tilde b^J~\equiv~b^J-M^J_Kc^K~=~0,~~~~~~~~
M^J_K~\equiv~\left.\lam^J_L~\tilde\mu^L_K \right|_{\T^I~=~0}.
\ee
We define a reduced subspace (denoted with $\star$) by $\T^I=\tb^I=0$.
It is closed since $\D^2 \T^I~=~0$ due to \bref{ancond}.

In order to reduce these variables explicitly we make a canonical
transformation generated by 
\be
{W}~=~\tb^*_J(b^J-M^J_Lc^L)~+~\tilde\phi^*_j\phi^j ~+~
\tilde c^*_\mu c^\mu~+~\tilde\T^*_\mu\T^\mu~+~\tilde b^*_\A b^\A.
\ee
It defines new field
\be
\tb^J~=~b^J-M^J_Lc^L
\ee
and anti-fields by
\be
c^*_L~=~\tilde c^*_L-\tilde b^*_J M^J_L,~~~~~
\phi^*_j~=~\tilde \phi^*_j-\tilde b^*_J M^J_{L,j} c^L,~~~~~
\T^*_\A~=~\tilde \T^*_\A-\tilde b^*_J M^J_{L,\A}c^L.
\ee
Other variables are unchanged.

The BRST transformation of functions in reduced space is generated by
$S^\star$,
\be
\left.\D f(\tilde\Phi,\tilde\Phi^*)\right|_{\star}~=~
\left.( f(\tilde\Phi,\tilde\Phi^*),\tilde S)\right|_{\star}~=~
( f(\tilde\Phi,\tilde\Phi^*), S^\star),~~~~~
S^\star~\equiv~\left.\tilde S\right|_{\star}.
\ee
It is also shown that
\be
\left.(\tilde S,\tilde S)\right|_{\star}~=~(S^\star,S^\star).
\ee
Thus $S^\star$ plays the role of $\tilde S$ in the reduced space.
It is
\bea
S^\star~=~S_0(\tilde\phi)~+~\tilde\phi^*_j~R^j_\mu(\tilde\phi)~\tilde c^\mu~+~
\frac12\tilde c^*_\mu~T^\mu_{\rho\s}(\tilde \phi)~\tilde c^\s \tilde c^\rho~+~
\tilde \T^*_\A~\{-{\tilde{\mu'}}^{\A}_\nu~\tilde c^\nu~+~
\xi^\A_A~\tilde b^A\}~+~
\nn\\
\frac12 \tilde b^*_D \{ T^D_{AB}(F) \tilde b^B\tilde b^A~+~
2 ~ T^D_{IB}(F)\tilde b^BM^I_L \tilde c^L~+~
 T^D_{IJ}(F)M^J_K \tilde c^KM^I_L \tilde c^L\}. 
\label{extaction11}
\eea
Note it does not depend on $\tilde b^*_J$ and the transformation of $\T^\A$
has been changed as
\be
\D\tilde \T^\A~=~\{-{\tilde{\mu'}}^{\A}_\nu~\tilde c^\nu~
+~\xi^\A_A~\tilde b^A\},~~~~~~~~~~
{\tilde{\mu'}}^{\A}_I~\equiv~{\tilde{\mu}}^{\A}_I-\mu^\A_JM^J_I,~~~
{\tilde{\mu}}^{\A'}_\B~\equiv~{\tilde{\mu}}^{\A}_\B.
\label{delTalpredA}
\ee
$S^\star$ also satisfies CME only on shell of the WZ equation of motion 
\be
(S^\star,~S^\star)~=~\left.(\tilde S,\tilde S)\right|_\star~=~
-\{\T^*_\A\mu^\A_\nu Z^\nu_a(F)~+~\tilde b^*_A~T^A_{a\tB}(F)~b^\tB\}~
T^a_{\tD\tE}(F)b^\tE b^\tD,
\ee
where $b^\A~=~\tilde b^\A$ and $b^I~=~M^I_J \tilde c^J$.


\end{document}